\documentclass[
 amsmath,amssymb,
 aps,
prl,
twocolumn,
]{revtex4-2}
\usepackage[colorlinks,linkcolor=blue,anchorcolor=blue,citecolor=blue,urlcolor=blue]{hyperref}
\usepackage[colorlinks,linkcolor=blue]{hyperref}
\usepackage{graphicx}
\usepackage{epstopdf}
\usepackage{dcolumn}
\usepackage{bm}
\usepackage{hyperref}
\usepackage[mathlines]{lineno}
\usepackage{multirow}

\begin{document}

\preprint{APS/123-QED}

\title{Phenomenological theory in reentrant uranium-based superconductors}

\author{Xilin Feng$^{1,2}$,Qiang Zhang$^{1}$,Jiangping Hu$^{1,3}$}
\email{jphu@iphy.ac.cn}\label{*}

\affiliation{%
 $^1$Beijing National Laboratory for Condensed Matter Physics and Institute of Physics, Chinese Academy of Sciences, Beijing 100190, China
 \\$^2$School of Physical Sciences, University of Chinese Academy of Sciences, Beijing 100190, China
 \\$^3$CAS Center of Excellence in Topological Quantum Computation and Kavli Institute of Theoretical Sciences,
	University of Chinese Academy of Sciences, Beijing 100190, China
}
\date{19 March 2019}

\begin{abstract}
We develop a phenomenological theory for the family of uranium-based heavy fermion superconductors ($URhGe$, $UCoGe$, and  $UTe_2$ ). The theory unifies the understanding of both superconductivity(SC)  with a weak magnetic field and reentrant superconductivity(RSC) that appears near the first-order transition line with a  high magnetic field. It is shown that  the magnetizations along the easy and hard axis have opposite effects on SC.  The RSC is induced by the fluctuation parallel to the direction of the magnetic field. The theory makes specific predictions about   the variation of  triplet SC order parameters $\vec{d}$  with applied external magnetic fields and the existence of a metastable state for the appearance of the RSC.
\end{abstract}
\maketitle
Heavy-fermion superconductors $UCoGe$, and $URhGe$ are promising spin triplet superconductors. The spin triplet pairing is supported by their highly anisotropic upper critical fields which greatly exceed the Pauli limit along all three crystallographic directions \cite{exp3,exp4,exp6,exp7,exp13}, and the coexistence of ferromagnetism (FM) and superconductivity (SC) \cite{exp8,exp9,exp10,exp11}.

Very recently, another uranium-based superconductor(UBS) $UTe_2$ has been found. Considerable researches have been conducted, such as a large residual Sommerfeld coefficient\cite{exp1,exp2}, coexistence of ferromagnetic fluctuations and superconductivity\cite{exp12,exp14}, field-boosted superconductivity\cite{exp18,exp19}, chiral superconducting state\cite{nature579}, quasi-two-dimensional Fermi surface\cite{prlxu} and so on. The new superconductor shares many common features with the previous counterparts, such as highly anisotropic upper critical fields and reentrant superconductivity (RSC)  under high magnetic fields. However, unlike the previous ones,  there is no sign of FM order in $UTe_2$ down to 25 mK \cite{exp12,exp14}. In all these superconductors, the SC transition temperature,  $T_{c}$, is first suppressed by the magnetic field ($h_y$) perpendicular to both the hardest ($x$) and easy axis ($z$). But  when the magnetic field   is strong enough, the $T_{c}$ arises again \cite{exp15,exp16,exp17,exp18,exp19,exp+1,exp+2}.

The difference between these superconductors brings new challenges and calls for a unified understanding.  On the basis of Landau phenomenological theory and weak-coupling theory for $URhGe$ given by  Mineev \cite{they1,they2},  the jump of the magnetic moment $m_{z0}$ enhances the fluctuations along the easy axis to  induce the RSC. This mechanism can not be applied to understand  the RSC in $UTe_2$\cite{exp18,exp19} because $UTe_2$ has no magnetic order along the easy-axis \cite{exp13,exp14,exp18,exp19}.    The increase of the fluctuation along the easy axis cannot be the only cause of  the RSC. Experimentally,  it has also been found that both the longitudinal (along the easy axis) and  transverse (along the magnetic field) fluctuations  exist near the RSC region in $URh_{0.9}Co_{0.1}Ge$ by $^{59}Co$ nuclear magnetic resonance (NMR) measurements \cite{exp20}.

Herein, we  generalize the phenomenological theory of the spin fluctuation feedback effect (SFFE) proposed by Amin et.al \cite{they5} to explain  the physics in the family of UBS.  We show that the decrease of $T_{c}$ in a weak magnetic field and the appearance (disappearance) of the RSC near the first-order transition in $URhGe$, $UCoGe$, and $UTe_2$ can be understood in a unified manner.  In the weak magnetic field region,  $T_{c}$ decreases with the decrease of static magnetic order along easy axis  and the increase of magnetic moment along field directions. In the strong field region, the RSC is caused by the fluctuations along magnetic field directions. However, RSC can be killed by destroying the metastable state near a  first order transition and a sudden increase of magnetic moment along the field directions.  Our theory further  predicts the $\vec{d}$ vector of the RSC in these superconductors and the metastable RSC state during the magnetic-hysteresis-loop, providing a sound theoretical basis for further investigation of the RSC in a microscopic theory.

\textit{Ferromagnetic SC}-we first focus on the SC and RSC in FM UBS, and take $URhGe$  as an example.  The phase diagram is sketched  in Fig.\ref{fig:1}.  With weak magnetic fields, the  SC  coexists with  FM, and as the spin-orbital coupling is strong, the symmetry is described by the magnetic group $D_{2h}(E,C_{2z},I,\sigma_{xy})$\cite{fujimori2016}. The spin triplet SC order parameter, $\vec{d}$ vector, is expanded in the basis of the $A_u$ or $B_u$ anti-symmetric co-representation of this magnetic group. For both $A_u$ and $B_u$, the free energy of magnetic ($\vec{m}$) and magnetism-SC coupling parts are the invariants \cite{Sigrist-RMP91, QZ-20}  of the magnetic group \cite{SI1}:
\begin{equation}
\label{equ:fsc-m1}
\begin{split}
f_{sc-m}=A_{1i}(m_i)^2+B_{ij}(m_i)^2(m_j)^2-h_{y}m_{y}\\+K_{1ij}(m_i)^2|d_j|^2+{K_{2z}}m_z(i\vec{d}\times\vec{d}^*)_z,
\end{split}
\end{equation}
here $i,j=x,y,z$ and  the repeated subscripts indicate summation throughout the paper. Except $A_{1z}<0$, other $A_{1i},B_{ij}, K_{1ij}$'s are positive or positive-definite to ensure the FM ground state. Positive $ K_{2i}$ are the amplitudes of the couplings between the FM and SC order parameters \cite{they5}.
\begin{figure}
    \centering
    \includegraphics[width=1\linewidth]{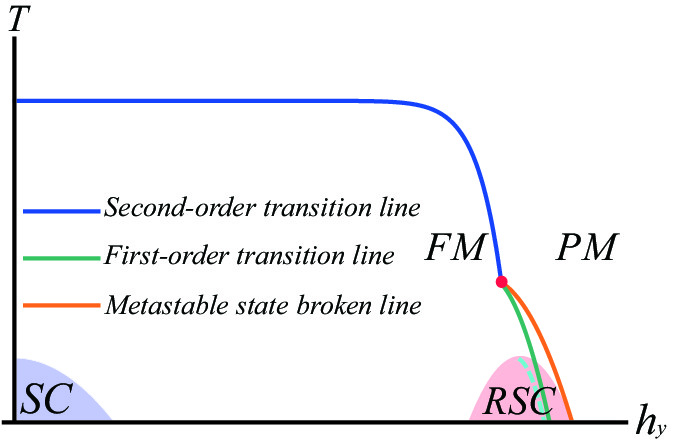}
    \caption{The sketched phase diagram of $URhGe$. The broken cyan line indicates the proposed  magnetic-hysteresis-loop type of behavior for  the upper critical magnetic field for RSC.}
    \label{fig:1}
\end{figure}

In the weak magnetic field region, the relevant magnetic part of the   free energy can be simplified as
    $f_{m,1}\approx A_{1z}m_{z}^2+B_{z}m_{z}^4+A_{1y}m_{y}^2-h_{y}m_{y}$. The minimization  gives the magnetic moment: $\vec{m_{0}}=(0,\frac{h_{y}}{2{A_{1y}}}, \sqrt{\frac{-A_{1z}}{2B_{z}}})$.  We integrate out the magnetic fluctuation $\delta\vec{m}$ ( $\vec{m}=\vec{m_{0}}+\delta \vec{m}$) in $f_{sc-m}$  to get the effective SC free energy:
\begin{equation}\label{eq:fsceff1}
    \begin{split}
f_{sc}=\alpha_{i}'|d_{i}|^2+K_{2z}m_{z0}p_{z}+\beta_{1ij}|d_{i}|^2|d_{j}|^2+\beta_{2z}p_{z}^2\\+\beta_{3iz}\cdot p_{z} |d_{i}|^2,
    \end{split}
\end{equation}
where $\vec{p}=i \vec{d}\times (\vec{d})^*$ and  $\alpha_{i}'= \alpha_{i}+K_{1zi}m_{z0}^2+\frac{K_{1zi}}{8B_{z}m_{z0}^2}+\frac{K_{1yi}}{2A_{1y}}+K_{1yi}m_{y0}^2$, with  $\alpha_{i}$  being the bare quadratic coefficient of SC without SFFE. The positive or positive-definite  quartic coefficients $\beta$'s are renormalized from the SFFE as listed in section.II of the SM \cite{SI2}. Here different components $d_i$ are not degenerate.

To track the evolution of $T_c$ under varying magnetic field, we rescale the $\vec{d}$ vector with $\alpha'_i|d_i|^2=\alpha'|d^I|^2$ \cite{SI2}, where $\alpha'$ is the   $\alpha'_i $ corresponding to the highest $T_c$. By minimizing the free energy,   we obtain the non-unitary SC with rescaled order parameter $\vec{d}^{I}=\frac{d_{0}}{\sqrt{2}}(\vec{r})(1,-i,0)$\cite{SI3}. This SC state has intrinsic $z$-polarized magnetic moment proportional to $\vec{p}^{I}=-d_{0}^2({\vec{r}})\hat{z}$ \cite{Sigrist-RMP91} and  $T_c=\Delta T_c+T_{c0}$, with :
\begin{equation}\label{tc1}
\Delta T_{c}=-\frac{K_{1yi}m_{y0}^2+K_{1zi}m_{z0}^2-K'_{2z}m_{z0}}{\alpha_0} -\frac{K_{1zi}}{8B_{z}m_{z0}^2\alpha_{0}},
\end{equation}
here   $\alpha_i=\alpha_0(T-T_{c0})$ with $\alpha'_i=\alpha'$ and $K'_{2z}=\frac{\sqrt{\alpha'_x\alpha'_y}}{\alpha'}K_{2z}$ (here the weak temperature dependence of the $K_{2z}'$ can be ignored as discussed in SM \cite{SI2}).  For a weak FM superconductor \cite{they2,exp24}, we can assume that  $m_{z0}^2<\frac{1}{2\sqrt{2B_{z}}}$. In this case, it can be seen from Eq.\ref{tc1} that either the decrease of $m_{z0}$ or the increase of $m_{y0}$ results in the decrease of $T_{c}$.  Namely, $T_{c}$ decreases with increasing magnetic field $h_y$, corresponding to SC phase of $URhGe$ as shown in Fig.\ref{fig:1}. By the way, from the Eq.\ref{tc1}, we can see that the $K_{1yi}$ and $K_{1zi}$ coupling terms dominate when the magnetic field along y-axis is weak at least in $URhGe$, since the $K'_{2z}m_{z0}$ terms could not lead to the disappearance of the $T_{c}$.

Now we consider the strong magnetic field region to discuss the rotation of the $\vec{d}$ vector and appearance of the RSC close to the magnetic first-order transition. When the magnetic field is strong enough, the symmetry of the $URhGe$ and $UCoGe$ is described by the magnetic group  $D_{2h}(E,C_{2y},I,\sigma_{xz})$. We obtain \cite{SI1}:
\begin{equation}\label{equ:fsc-m1A2}
\begin{split}
f_{sc-m}=A_{1i}(m_i)^2+B_{ij}(m_i)^2(m_j)^2-h_{y}m_{y}\\+K_{1ij}(m_i)^2|d_{j}|^2+{K_{2y}}m_{y}(i\vec{d}\times\vec{d}^*)_{y}-\lambda h_{y}p_{y}.
\end{split}
\end{equation}
On the low field side of the first-order transition,  the y-component of the magnetic moment enters the free energy as:
$f_{m,2}\approx f_{m,1}+B_{yz}m^2_ym^2_z$,
giving the magnetic moments as $m_{z0}^2=-\frac{A_{1z}+B_{yz}m_{y0}^2}{2B_{z}}$ and $m_{y0}=\frac{h_y}{2(A_{1y}+B_{yz}m_{z0}^2)}$. Following the same procedure, integrating out magnetic fluctuations and rescaling the $\vec{d}$ vector, we obtain the effective free energy:
\begin{equation}\label{fsec1}
f'_{sc}\approx \alpha'|\vec{d}^{I}|^2+K_{2y}'m_{y0}p_{y}^{I}-\lambda' h_{y}p_{y}^{I}+HO,
\end{equation}
here again $\alpha'=\alpha'_{i}$ and the high order ($HO$) terms are not specified. The  coupling $K'_{2y}$ makes the SC with $\vec{d}^{I}=\frac{d_{0}(\vec{r})}{\sqrt{2}}(1,0,\pm i)$ and $\vec{p}^{I}=\mp d_0(\vec{r})^2\hat{y}$ at the highest $T_{c}$. Close to the first-order transition critical magnetic field $h_{m}$, before the sudden jump of $m_{z0}$, the variation of $m_{z0} $ is small, so $\Delta T_c$ tuned by the magnetic field can be approximated as:
\begin{equation}\label{tc-c1}
\Delta T_{c}=-\frac{1}{\alpha_0}(\frac{K_{1yi}}{4A^{'2}_{1y}}h_{y}^2-|\frac{K_{2y}'}{2A'_{1y}}-\lambda'|h_{y}),
\end{equation}
where $A_{1y}^{'}=A_{1y}+B_{yz}m_{z0}^{2}$, $K'_{2y}=\frac{\sqrt{\alpha'_x\alpha'_z}}{\alpha'}K_{2y}$, $\lambda'=\frac{\sqrt{\alpha'_x\alpha'_z}}{\alpha'}\lambda$. The phase diagram in the strong magnetic field region can be explained if we assume $h_{m}<h_{q}\equiv \frac{|{K_{2y}'}{A'_{1y}}-2\lambda'{A'^2_{1y}}|}{{K_{1yi}}}$.In this case, from Eq.\ref{tc-c1}, the $T_{c}$  increases with increasing magnetic field ($h_{y}>0$) at first and then decrease when the magnetic filed $h_{y}>h_{q}$. However, it is noteworthy that Eq.\ref{tc-c1} is valid only for the region which is on the left side of the first-order transition line and close to it. From Eq.\ref{tc-c1}, we know that the key coupling terms which leads to the appearance of the reentrant superconductivity under strong magnetic field are $K_{2y}$ and $\lambda$ terms. When the magnetic field continues to increase and exceeds the critical value $h_{m}$, as will be analyzed next, the RSC disappears with increasing magnetic field.

The first order transition  and the disappearing of RSC close to it can be further understood within our theory. With a strong magnetic field and a small $m_{z0}$ in the FM phase,  the free energy to describe the first order transition can be derived \cite{they1,they2}:
\begin{equation}\label{fm2}
f_{m}=-\frac{h_y^2}{4A_{1y}}+A^{'}_{1z} m_z^2+B^{'}_{z}m_{z}^4+C^{'}_{z}m_{z}^6,
\end{equation}
where $A^{'}_{1z}=A_{1z}+\frac{B_{yz}h_{y}}{4A_{1y}^{2}}$, $B_{z}^{'}=B_{z}-\frac{B_{yz}^2h_{y}^2}{4A_{1y}^3}$, and $C^{'}_z=C_{z}+\frac{B_{yz}^3h_{y}^2}{4A_{1y}^4}$. So one can learn from Eq.\ref{fm2} that the magnetic field $h_{y}$ modifies the coefficients of the free energy $f_{m}$. Thus the magnetic moment dependence of the free energy  changes with increasing magnetic field as shown in Fig.\ref{fig:3}.
\begin{figure}
    \centering
    \includegraphics[width=1\linewidth]{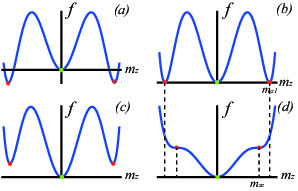}
    \caption{The $f_m$-$m_z$ relations from Eq.\ref{fm2} with  $B^{'}_z<0, A^{'}_{1z}>0, C^{'}_{z}>0$ for gradually increasing $h_y$: (a) FM state with $h_y=0$. (b) First order transition point $m_{z0}=m_{z1}$. (c)A metastable state $m_{ze}<m_{z0}<m_{z1}$. (d) The local minima broken state $m_{z0}=m_{ze}$.}
    \label{fig:3}
\end{figure}
From Eq.\ref{fm2}, we can derive \cite{SI4} the condition for the first-order transition, $m_{z1}^2=-\frac{B'_{z}}{2C'_{z}}$ as well as the condition that the local minima are broken: $m_{z e}^2=-\frac{B^{'}_{z}}{3C^{'}_{z}}$.  Here as $m_{z1}>m_{ze}$, there is a metastable state as displayed in Fig.\ref{fig:3} $(c)$, corresponding to the state between the green and orange lines in Fig.\ref{fig:1}. During the up-sweep of magnetic field, the system can cross the first-order transition line, the magnetic moment $m_{z0}$ does not collapse abruptly to zero but decreases continuously before the local minima are broken.

The existence of the metastable state is important to the RSC.  By substituting $\frac{\partial f_{m}}{\partial m_{z}}|_{m_{z0}}=0$ and ${m_{z}}={m_{z0}}+\delta {m_{z}}$ into the free energy Eq.\ref{fm2}, one can get the magnetic part of the free energy $f_{m}$.
Using this $f_m$ and the new $\vec{p}$ is parallel to the direction of the magnetic field which can be derived from $\vec{d}^{I}=\frac{d_{0}(\vec{r})}{\sqrt{2}}(1,0,-i)$,
we obtain $\Delta T_c$ from $f_{sc-m}$ with the same method as before:
\begin{align}\label{tc2}
\Delta T_{c}=-\frac{K_{1zi}}{2(6C^{'}_z m_{z0}^4-2A^{'}_{1z})\alpha_{0}}-\frac{K_{1zi}m_{z0}^2}{\alpha_{0}}+\frac{|\lambda'| h_y}{\alpha_{0}}.
\end{align}
The second term in Eq.\ref{tc2} shows that if the metastable state is broken, namely, $m_{z0}=m_{ze}$ ($m_{ze}^4= \frac{A_{1z}'}{3C_{z}^{'}}$, as shown in Fig.\ref{fig:3} $(d)$), $T_{c}$  reaches  $-\infty$, indicating  the truncation of RSC right before the metastable state broken line.  Moreover, the Eq.\ref{tc2} also shows the $K_{1zi}$ coupling terms are the key coupling which are responsible for the truncation of the RSC.

\textit{Paramagnetic SC}-there are several known experimental facts for the paramagnetic UBS, $UTe_2$\cite{exp13,exp14}.  The SC as the weak field region is initially  suppressed by the increasing magnetic field $h_y$. However, when the magnetic field is sufficiently strong, the RSC appears. Finally, when the magnetic field arrives at 34.9T \cite{exp25}, a first order transition occurs with the increasing jump of the magnetic moment $m_{y0}$, and the RSC disappears simultaneously. The phase diagram of the $UTe_2$ is summarized in Fig.\ref{fig:2}.
\begin{figure}
    \centering
    \includegraphics[width=1\linewidth]{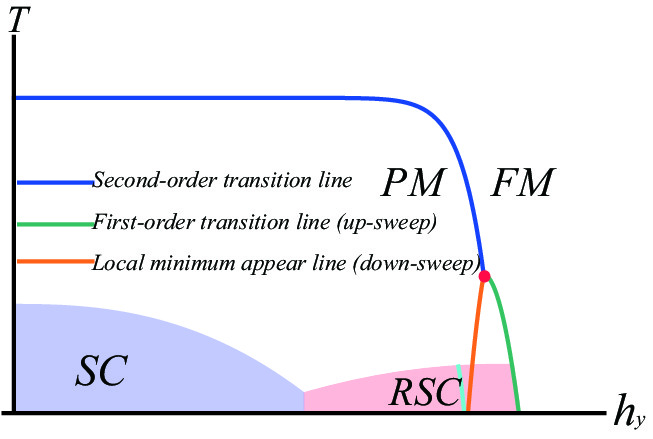}
    \caption{The sketched phase diagram of $UTe_2$. The solid cyan line indicate the magnetic-hysteresis-loop type of behavior of the upper critical field for RSC.}
    \label{fig:2}
\end{figure}

Due to the absence of FM order, the absence of the $K_{2i}$ coupling terms in our free energy hardly supports the non-unitary triplet SC states in the weak magnetic field region,  consistent with the measurements of heat capacity and thermal conductivity in $UTe_2$ which indicates the point-node gap structure \cite{exp+3}.
Similar to the method in the ferromagnetic UBS, $\Delta T_c$ can be derived:
\begin{equation}\label{tc-3a1}
\Delta T_{c}= -(\frac{K_{1zi}}{2A_{1z}}+\frac{K_{1yi}}{2A_{1y}})-K_{1yi}m_{y0}^2,
\end{equation}
where $T_{c0}$ is the superconducting critical temperature without the SFFE. Since $m_{y0}$  increases with increasing $h_{y}$, the Eq.\ref{tc-3a1} implies $T_{c}$  decrease as shown in Fig.\ref{fig:2}.  In addition, from the Eq.\ref{tc-3a1}, we can learn that: for $UTe_{2}$, the key coupling terms in the weak magnetic field region are $K_{1yi}$ coupling terms.

However, when the magnetic field is strong enough, from our theory, the symmetry of $UTe_{2}$ can be described by a magnetic group $D_{2h}(E,C_{2y},I,\sigma_{xz})$, thus the non-unitary SC can appear because of the $K_{2i}$ coupling term in Eq.\ref{equ:fsc-m1A2} as discussed before in FM UBS. For both co-representations $A_{u}$ and $B_{u}$, the free energy can be expressed as \cite{SI1}:
\begin{equation}\label{fscm3b1}
\begin{split}
f_{sc-m}=A_{1z}m_{z}^2+A_{1y}m_{y}^2+K_{1zi}m_{z}^2|d_{i}|^2+K_{1yj}m_{y}^2|d_{j}|^2\\+K_{2y}m_{y}p_{y}-\lambda h_{y}p_{y},
\end{split}
\end{equation}
 and $\Delta T_c$ can be derived as:
\begin{align}\label{sc-pmsc2}
\Delta T_{c}= -\frac{K_{1yi}m_{y0}^2}{\alpha _{0}}+\frac{|K_{2y}'-2\lambda' A_{1y}|}{\alpha _{0}}m_{y0}.
\end{align}
This parabolic function on $m_{y0}$ explains the RSC close to the first order transition line in $UTe_2$  and shows the key coupling terms which lead to the appearance of the RSC are $K_{2y}$ and $\lambda$ coupling terms.

Similar to the FM case, we can also describe the first-order transition and metastable state in $UTe_{2}$, which have been detected in experiment \cite{exp25}. In this case, the magnetic part of the free energy in a strong magnetic field can be written as
\begin{equation}\label{3cfm}
f_{m}=a_{y}m_{y}^2-c_{y}m_{y}^3+\frac{1}{2}b_{y}m_{y}^4-\mu_{1} h_{y} m_{y}+a_{z}m_{z}^2.
\end{equation}
 We can derive \cite{SI5} the condition for the first-order transition, $h_{yc1}=\frac{c_{y}}{\mu_{1}}(\frac{a_{y}}{b_{y}}-\frac{c^{2}_{y}}{2b^2_{y}})$ as well as the condition that the local minimum appears, $h_{yc2}=\frac{\Delta_{y} (8 a_{y} b_{y}-c_{y}\Delta_{y})}{36 b^2_{y} \mu_{1}}$, where $\Delta_{y}=3 c_{y}+\sqrt{9 c_{y}^2-12 a_{y} b_{y}}$. Here $h_{yc2}<h_{yc1}$, so the metastable state is located in the paramagnetic region as shown in Fig.\ref{fig:2}. Then considering the fluctuation $\vec{m}=(0,m_{y0}+\delta m_{y},\delta m_{z})$, we obtain
the total free energy  and $\Delta T_c$ as follow:
\begin{eqnarray}\label{3cfscm}
& &f_{sc-m}=a_{z}\delta m_{z}^2+(-a_{y}+\mu_{1} \frac{h_{y}}{m_{y0}})\delta m_{y}^2+K_{1zi}m_{z}^2|d_{i}|^2\nonumber \\
& &+K_{1yj}m_{y}^2|d_{j}|^2+K_{2y}m_{y} p_{y} -\lambda h_{y} p_{y},
\end{eqnarray}
\begin{equation}\label{3ctc}
\Delta T_{c}=-\frac{1}{\alpha_0}(\frac{K_{1yi}}{2(\mu_{1} \frac{h_{y}}{m_{y0}}-a_{y})}-K_{1yi}m_{y0}^2+|K_{2y}'m_{y0}-\lambda' h_{y}|).
\end{equation}
When the first-order transition happens, the magnetic moment $m_{y0}$ increases abruptly. The second term in Eq.\ref{3ctc} increases suddenly. As for the last two terms of Eq.\ref{3ctc},  the jump of the magnetic moment $m_{y0}$ can lead to $m_{y0}^{'}>>\frac{|K_{2y}'-2\lambda' A_{1y}|}{2K_{1yi}}$ which belongs to the right side of the first-order transition line.  In this case, the $T_{c}$  decrease abruptly as  shown in Fig.\ref{fig:2}. This explains the experimental observation of the sudden truncation of the RSC in $UTe_2$ upon the first order transition \cite{exp18,exp19} and shows the key coupling terms leading to this sudden truncation are $K_{1yi}$ coupling terms.

\textit{Summary}-we develop a phenomenological theory with respect to the full magnetic groups to describe the SC and RSC in UBS unifiedly. The theory explains the global phase diagram of this family of superconductors.  In our theory, the SC at weak magnetic region are suppressed with the increasing transverse magnetic field $h_y$, due to the energy cost from the mismatch of the induced transverse magnetic moment $m_{y0}$ with the $z$-polarized non-unitary SC order $p_z$ and the unitary SC order, for the ferromagnetic and paramagnetic superconductors, respectively.
However, the RSC in both ferromagnetic and paramagnetic superconductors are induced by the fluctuation parallel to the magnetic fields, rather than the sudden jump of the magnetic moment upon the first order transition.  Instead, the sudden jump of the magnetic moment indeed truncates the RSC and there should be a shift of the RSC dome upon a magnetic-hysteresis-loop type of measurement.

Moreover, due to the non-degenerate nature of the triplet SC $\vec{d}$ vector under the magnetic group, another interesting phenomenon of the multi-jump of the specific heat at different temperatures corresponding to the transition of each component might be observed.    The non-unitary coupling $K_{2z}$ term can further cause the splitting of the transition temperature of $d_x$ and $d_y$ from that of $d_z$, as derived with simplification in sec.VI of SM\cite{SI6}. Assuming small difference among the bare quadratic coefficients, their renormalization  would only tune  $T_c$, leaving the jumps of specific heat at the transition temperature $\frac{\Delta C}{T_c}$ intact to the varying magnetic field.

Our theory makes a few explicit  predictions. First,  we predict that the rotation of the spin-triplet pairing $\vec{d}$ vector in different magnetic field regions.  In ferromagnetic superconductors $UCoGe$ and $URhGe$, with increasing magnetic field, the rescaled $\vec{d}$ vector rotates from $\frac{d_{0}(\vec{r})}{\sqrt{2}}(1,-i,0)$ to $\frac{d_{0}(\vec{r})}{\sqrt{2}}(1,0,\pm i)$. In $UTe_{2}$, the SC is unitary at first. However with a high enough magnetic field,  it becomes a non-unitary SC with a rescaled $\vec{d}$ vector,  $\frac{d_{0}(\vec{r})}{\sqrt{2}}(1,0,\pm i)$. The rotation of the $\vec{d}$ vector by magnetic field was studied in Sr$_2$RuO$_4$\cite{Annett08PRB}  whose spin-triplet pairing symmetry has been  questioned\cite{brown19}. In principle, this prediction can be  examined experimentally in superconducting junctions made by these materials. The $\vec{d}$ vector can  be visualized from quasi-particle interference technique in STM experiments\cite{QHWang18PRB}.

Second, we predict  that it is the metastable state that ensures the extension of the RSC over  the right side of the first-order transition line in $URhGe$ and $UCoGe$. This prediction can be checked by performing  a magnetic-hysteresis-loop type measurement around the first order transition line. We can apply  a strong magnetic field to destroy the metastable state at first and then reduce it to induce the RSC.   The maximum of the upper critical magnetic field is predicted to have a magnetic-hysteresis-loop type of behavior. Namely, it is much smaller than  the one  with a normal procedure that the field crosses the first-order transition line from its left side. The RSC dome upon down-sweep magnetic field would shift to the left of the first order transition line as depicted by the broken cyan line in Fig.\ref{fig:1}.

Finally, we predict that the metastable state also exists in $UTe_2$ and affects the behavior of the RSC in $UTe_2$ because of the magnetic hysteresis \cite{exp25}. The metastable state indicates the remaining large magnetic moment $m_{y0}$ during the down-sweep process. Since the RSC is truncated by the sudden increase of $m_{y0}$, during the up-sweep, the RSC would exist until the first order transition line. However, during the   down-sweep, the magnetic moment does not decrease abruptly when the system crosses the first-order transition line  so that the RSC will not appear untill the system cross the metalstable state broken line (the cyan down-sweep line in Fig.\ref{fig:2}). (note: After we completed this paper, we notice that the magnetic-hysteresis-loop type of behavior  near the first-order transition line in $UTe_2$ were detected\cite{lin2020tuning}, which is a strong support for our theory.) By the way, we also notice that in $UTe_2$, a new reentrant superconductivity which exists only in the FM region has been detected in Ref.\cite{exp19}. In the frame of our theory, the reason why this reentrant superconductivity only exist in FM region is highly related with the down-sweep path in $URhGe$ and $UCoGe$, in which the superconductivity does not appear until the system cross the first-order transition line (as mentioned in our second prediction about $URhGe$ and $UCoGe$). In addition, Since the direction of the magnetic field is in a specific region between b-axis and c-axis, it may also be related with both of the field-induced fluctuations along c-axis and b-axis, which need further investigations.

{\it Acknowledgement} We thank J Singleton and HQ Yuan for helpful discussions. Q.Zhang acknowledges the support from the International Young Scientist Fellowship of Institute of Physics CAS (Grant No. 2017002) and the Postdoctoral International Program from China Postdoctoral Science Foundation (Grant No. Y8BK131T61).  The work is supported by the
Ministry of Science and Technology of China 973 program
(No. 2017YFA0303100), National
Science Foundation of China (Grant No. NSFC11888101), and the
Strategic Priority Research Program of CAS (Grant
No.XDB28000000).
\bibliography{v4}

\newpage

\begin{appendix}
\renewcommand{\theequation}{A.\arabic{equation}}
\renewcommand{\thefigure}{AF.\arabic{figure}}
\renewcommand{\thetable}{AT.\arabic{table}}

\maketitle

\section{I. magnetic group and invariants}
When spin-orbital coupling is strong and the magnetic field along $y$-direction is weak, the symmetry of the $URhGe$ and $UCoGe$ with ferromagnetic moment along $z$-direction is described by $D_{2h}(E,C_{2z},I,\sigma_{xy})$\cite{fujimori2016}. We list the changes of components of magnetic moment when elements belonging to magnetic group $D_{2h}(E,C_{2z},I,\sigma_{xy})$ act on them in Table.\ref{tab+1}:

\begin{table}[ht]
\caption{Changes of the components of the magnetic moment when spin-orbital coupling is strong.}\label{tab+1}
\begin{ruledtabular}
\begin{tabular}{c|cccccccc}

components & $E$                                    & $C_{2z}$                                 & $RC_{2x}$                              & $RC_{2y}$  &I&$\sigma_{xy}$ &R$\sigma_{yz}$ &R$ \sigma_{xz} $                            \\ \hline
$m_{x}$    & $m_{x}$ & $-m_{x}$ & $-m_{x}$ & $m_{x}$ & $m_{x}$ & $-m_{x}$ & $-m_{x}$ & $m_{x}$  \\
$m_{y}$    & $m_{y}$ & $-m_{y}$ & $m_{y}$ & $-m_{y}$ & $m_{y}$ & $-m_{y}$ & $m_{y}$ & $-m_{y}$ \\
$m_{z}$    & $m_{z}$ & $m_{z}$ & $m_{z}$ & $m_{z}$ & $m_{z}$ & $m_{z}$ & $m_{z}$ & $m_{z}$\\
\end{tabular}
\end{ruledtabular}
\end{table}

There are two nonequivalent irreducible anti-symmetric  co-representations $A_{u}$ and $B_{u}$ in this magnetic group induced from the one-dimensional representations of the unitary invariant subgroup $(E, C_{2z},I,\sigma_{xy})$. We list these two anti-symmetric co-representations and their basis in Table.\ref{tab1}, because we only care about the spin triplet superconductivity: ( Notice: $\vec{e}_{i}$ are axial vectors, which are invariant under inversion $I$)

\begin{table}[ht]
\caption{The anti-symmetric co-representation of $D_{2h}(E,C_{2z},I,\sigma_{xy})$ and the basises of them
respectively}\label{tab1}
\begin{ruledtabular}
\begin{tabular}{c|ccccccccc}
Co-Rep & $E$                                    & $C_{2z}$                                 & $RC_{2x}$                              & $RC_{2y}$   &I&$\sigma_{xy}$ &$R\sigma_{yz}$&$\sigma_{xz}$  & basis                          \\ \hline
$A_{u}$    & 1 & 1 & 1 & 1 & -1 & -1 & -1 & -1 & $\vec{d}_{A}$ \\
$B_{u}$    & 1 & -1 & -1 & 1 & -1 & 1 & 1 & -1& $\vec{d}_{B}$ \\
\end{tabular}
\end{ruledtabular}
\end{table}

In this table, the representation basis are:
\begin{equation}\label{basis-A}
    \vec{d}_{A}=\frac{1}{2}[(\eta_{1}{x}-i\eta_{2}{y})(\vec{e}_{x}+i\vec{e}_{y})+(\eta_{3}{x}+i\eta_{4}{y})(\vec{e}_{x}-i\vec{e}_{y})]+\eta_{5}{z}\vec{e}_{z},
\end{equation}
\begin{equation}\label{basis-B}
\begin{split}
    \vec{d}_{B}=\frac{1}{2}[{z}\zeta_{3}(\vec{e}_{x}+i\vec{e}_{y})+{z}\zeta_{4}(\vec{e}_{x}-i\vec{e}_{y})]\\+(\zeta_{1}x+i\zeta_{2}y)\vec{e}_{z},
    \end{split}
\end{equation}
with $\eta_\mu$'s and $\zeta_\mu$'s being real constants (or owning the same phase). Each $d_i\vec{e}_i$ is a basis of that representation and they are independent and thus non-degenerate. The time reversal symmetry broken can be expressed by the relative phase of the components of the $\vec{d}$ vectors. Let $\phi_{Az}$ and $\phi_{Bx}$ being zeros, then
\begin{align}
    \phi_{Ax}&=arctan(\frac{(\eta_{4}-\eta_{2})y}{(\eta_{1}+\eta_{3})x}),\\
    \phi_{Ay}&=arctan(\frac{(\eta_{1}-\eta_{3})x}{(\eta_{2}+\eta_{4})y}),\\
	\phi_{By}&=\frac{\pi}{2}sign({\zeta_{3}-\zeta_{4}}),\\
	\phi_{Bz}&=arctan(\frac{\zeta_{2}y}{\zeta_{1}x})\not=0.
\end{align}

Then we turn to the the terms coupling magnetic moment and superconductivity in free energy $f_{sc-m}$ (the Eq.(1) in the main text). All terms in free energy has to be invariant  \cite{Sigrist-RMP91, QZ-20} under all operations belonging to magnetic group $D_{2h}(E,C_{2z},I,\sigma_{xy})$ and $U(1)$ gauge and we list them in detail below.

For the $\vec{d}$ belonging to the co-representation $A_{u}$, the quadratic terms  from Eq.\ref{basis-A} are:
\begin{equation}\label{dA1}
\begin{split}
    d_{Ax}^2=\frac{1}{4}[x^2(\eta_{1}+\eta_{3})^2+y^2(\eta_{4}-\eta_{2})^2]\\
    d_{Ay}^2=\frac{1}{4}[x^2(\eta_{1}-\eta_{3})^2+y^2(\eta_{2}+\eta_{4})^2]\\
    d_{Az}^2=(\eta_{5})^2z^2.
    \end{split}
\end{equation}
They are invariant under all operations belonging to the magnetic group $D_{2h}(E,C_{2z},I,\sigma_{xy})$, so do $m^2_i$. Therefore the term $K_{1ij}m_{i}^2|d_{j}|^2$ can be included in the free energy. The term $K_{2i}m_{i}(i\vec{d}\times\vec{d}^*)_{i}, (i=x,y,z)$ can also be derived from Eq.\ref{basis-A}:
\begin{equation}\label{dA2}
\begin{split}
    x:-K_{2x}m_{x}zx\eta_{5}(\eta_{1}-\eta_{3})\\y:K_{2y}m_{y}zy\eta_{5}(\eta_{4}-\eta_{2})\\z:\frac{1}{2}m_{z}K_{2z}[x^2(\eta_{1}^2-\eta_{3}^2)-y^2(\eta_{4}^2-\eta_{2}^2)].
    \end{split}
\end{equation}
Terms in Eq.\ref{dA2} are all invariant under all operations belonging to the magnetic group $D_{2h}(E,C_{2z},I,\sigma_{xy})$. However, because the symmetry of these two materials $URhGe$ and $UCoGe$ is described by the magnetic group $D_{2h}(E,C_{2z},I,\sigma_{xy})$ when the magnetic field is weak, the $x$ and $y$ components of the spin average of the pairing state can't be finite. So only the term $ K_{2z}m_{z}(i \vec{d}\times \vec{d}^*)_{z}$ can be included in free energy $f_{sc-m}$ (Einstein summation convention is used here and below without special mention):
\begin{equation}
\begin{split}
f_{sc-m}=A_{1i}(m_i)^2+B_{ij}(m_i)^2(m_j)^2-h_{y}m_{y}\\+K_{1ij}(m_i)^2|d_{j}|^2+{K_{2z}}m_{z}(i\vec{d}\times\vec{d}^*)_{z},
\end{split}
\end{equation}
where $i,j=x,y,z$.

For the $\vec{d}$ belonging to the co-representation $B_{u}$, the quadratic terms   from Eq.\ref{basis-B} are:
\begin{equation}\label{dB1}
\begin{split}
    d_{Bx}^2=\frac{1}{4}z^2(\zeta_{3}+\zeta_{4})^2\\
    d_{By}^2=\frac{1}{4}z^2(\zeta_{3}-\zeta_{4})^2\\
    d_{Bz}^2=(\zeta_{1}^2 x^2+\zeta_{2}^2 y^2).
    \end{split}
\end{equation}
We can learn from the Eq.\ref{dB1} that $K_{1ij}m_{i}^2|d_{j}|^2$ can be included in the free energy, since it is invariant under all operations belonging to the magnetic group $D_{2h}(E,C_{2z},I,\sigma_{xy})$.

Moreover, we can derived the exact form of $\sum_{i}K_{2i}m_{i}(i\vec{d}\times\vec{d}^*)_{i}$ from Eq.\ref{dB1}:
\begin{equation}\label{dB2}
\begin{split}
    x:-K_{2x}m_{x}zx\zeta_{1}(\zeta_{3}-\zeta_{4})\\y:-K_{2y}m_{y}yz\zeta_{2}(\zeta_{3}+\zeta_{4})\\z:\frac{1}{2}K_{2z}m_{z}z^2(\zeta_{3}^2-\zeta_{4}^2).
    \end{split}
\end{equation}
In Eq.\ref{dB2}, all terms are invariant under all operations belonging to the magnetic group. As in co-representation $A_{u}$, only $K_{2z}m_{z}(i\vec{d}\times\vec{d}^*)_{z}$ can be included in the free energy, the free energy $f_{sc-m}$ of the co-representation $B_{u}$ is the same as the co-representation $A_{u}$.\\

When the magnetic filed along y-axis is strong enough, the magnetic group of $URhGe$ and $UCoGe$ is $D_{2h}(E,C_{2y},I,\sigma_{zx})$, there are also two co-representations in this magnetic group, as shown in Table.\ref{tab2}:
\begin{table}[ht]
\caption{The co-representations of $D_{2h}(E,C_{2y},I,\sigma_{xz})$ and the basises of them respectively}\label{tab2}
\begin{ruledtabular}
\begin{tabular}{c|ccccccccc}
Co-Rep & $E$                                    & $C_{2y}$                                 & $RC_{2z}$                              & $RC_{2x}$ &I&$\sigma_{zx}$&R$\sigma_{xy}$&R$\sigma_{yz}$    & basis                          \\ \hline
$A2_{u}$    & 1 & 1   & 1 & 1 & -1 & -1   & -1 & -1 & $\vec{d}_{A2}$ \\
$B2_{u}$    & 1 & -1 & -1 & 1 & -1 & 1 & 1 & -1& $\vec{d}_{B2}$ \\
\end{tabular}
\end{ruledtabular}
\end{table}

In this table:
\begin{equation}\label{basis-A2}
    \vec{d}_{A2}=\frac{1}{2}[(\eta_{1}{z}-i\eta_{2}{x})(\vec{e}_{z}+i\vec{e}_{x})+(\eta_{3}{z}+i\eta_{4}{x})(\vec{e}_{z}-i\vec{e}_{x})]+\eta_{5}{y}\vec{e}_{y},
\end{equation}
\begin{equation}\label{basis-B2}
\begin{split}
   \vec{d}_{B2}=\frac{1}{2}[{y}\zeta_{3}(\vec{e}_{z}+i\vec{e}_{x})+{y}\zeta_{4}(\vec{e}_{z}-i\vec{e}_{x})]\\+(\zeta_{1}z+i\zeta_{2}x)\vec{e}_{y}.
   \end{split}
\end{equation}
The same method as in weak magnetic region can be used here, then we can get the free energy $f_{sc-m}$ of co-representation $A2_{u}$ and $B2_{u}$ respectively. Here, we also consider the term coupling magnetic filed to superconductivity $-\lambda h_{y}(i\vec{d}\times\vec{d}^*)_{y}$.

For both co-representation $A2_{u}$ and $B2_{u}$:
\begin{equation}\label{fA2}
\begin{split}
f_{sc-m}=A_{1i}(m_i)^2+B_{ij}(m_i)^2(m_j)^2-h_{y}m_{y}\\+K_{1ij}(m_i)^2|d_{j}|^2+{K_{2y}}m_{y}(i\vec{d}\times\vec{d}^*)_{y}-\lambda h_{y}(i\vec{d}\times\vec{d}^*)_{y}.
\end{split}
\end{equation}

For $UTe_2$, in weak magnetic field region, the absence of the magnetic order leads to the absence of the $K_{2i}$ coupling terms. However, when the magnetic field is strong enough, the symmetry of the $UTe_{2}$ is described by the magnetic group $D_{2h}(E,C_{2y},I,\sigma_{xz})$, there are two  anti-symmetric co-representations $A3_{u}$, $B3_{u}$ and two symmetric co-representations $A3_{g}$, $B3_{g}$. Since we only care about the spin triplet superconductivity, we only care about the anti-symmetric co-representations.

\begin{table}[ht]
\caption{The co-representation of $D_{2h}(E,C_{2y},I,\sigma_{xz})$ and the basises of them
respectively}\label{tab3}
\begin{ruledtabular}
\begin{tabular}{c|cccccccccc}
Co-Rep & $E$                                    & $C_{2y}$                                 & $RC_{2z}$                              & $RC_{2x}$ & $I$ & $\sigma_{xz}$ &$R\sigma_{xy}$ &$R\sigma_{yz}$   & basis                          \\ \hline
$A3_u$    & 1 & 1   & 1 & 1 & -1 & -1 & -1 & -1 & $\vec{d}_{A3_u}$ \\
$B3_u$    & 1 & -1 & -1 & 1 & -1 & 1 & 1 & -1 & $\vec{d}_{B3_u}$ \\
\end{tabular}
\end{ruledtabular}
\end{table}
In table.\ref{tab3}, $\vec{d}_{A3_u}=\vec{d}_{A2}$, $\vec{d}_{B3_u}=\vec{d}_{B2}$ (Basises are the same as $URhGe$ and $UCoGe$ under a strong magnetic field along y-axis.). Thus the Eq.\ref{fA2} is also the free energy of both co-representations $A3_{u}$ and $B3_{u}$.

\section{II. Rescaling of the   $\vec{d}$ vector}
Here, we take $URhGe$ and $UCoGe$ in weak magnetic field region as an example. Starting from Eq.(1) in the main text, we integrate out the magnetic fluctuations $e^{-\int d^{4}x f_{sc}^{I}}=\int D(\delta \vec{m})e^{-\int d^{4}x f_{sc-m}}$ ($d^{4}x=d\tau d^{3}\vec{x}$, where $\tau$ is inverse temperature) to get the effective superconducting free energy, here the $f_{sc}^{I}$ is the modified part of the superconducting free energy.
The effective free energy can be expressed as:
\begin{equation}\label{fsc-1}
    \begin{split}
f_{sc}=(\alpha_{i}+K_{1zi}m_{z0}^2+\frac{K_{1zi}}{8B_{z}m_{z0}^2}+\frac{K_{1yi}}{2A_{1y}}+K_{1yi}m_{y0}^2)|d_{i}|^2\\+K_{2z}m_{z0}p_{z}+\beta_{1ij}|d_{i}|^2|d_{j}|^2+\beta_{2z}p_{z}^2+\beta_{3iz}\cdot p_{z} |d_{i}|^2,
    \end{split}
\end{equation}
where
\begin{center}
\begin{equation}\label{beta}
\begin{aligned}
\beta_{1ij}=\beta_{1ij0}-\frac{K_{1zi}K_{1zj}}{4B_{z}m_{z0}^2}-\frac{K_{1zi}K_{1zj}}{64B_{z}^2m_{z0}^2}-\frac{K_{1yi}K_{1yj}}{4A_{1y}^2}\\-\frac{K_{1yi}K_{1yj}m_{y0}^2}{A_{1y}},\\
\beta_{2z}=\beta_{2z0}-\frac{K_{2z}^2p_{z}^2}{16B_{z}m_{z0}^2},\\
\beta_{3iz}=\beta_{3i z0}-\frac{K_{1zi}K_{2z}}{4B_{z}m_{z0}}.
\end{aligned}
\end{equation}
\end{center}
Here, fluctuations are relatively small compared to $\beta_{1ij0}$, $\beta_{2z0}$ and $\beta_{3i z0}$. Thus the quartic terms in effective free energy are positive. As we can see in Eq.\ref{fsc-1}, the dependencies of these three $\alpha_{i}$ on magnetic moments have a similar form and the same trend. Moreover, although the quadratic terms in free energy can be described as: $\alpha_{i}|d_{i}|^2$, for each co-representation, there is only one highest $T_{c}$. To derive this $T_{c}$ conveniently, we will rescale the $\vec{d}$.

For each co-representation, the order parameter $\vec{d}$ can be written as a general form: $\vec{d}=d_{x}(\vec{r})\vec{e}_{x}+d_{y}(\vec{r})\vec{e}_{y}+d_{z}(\vec{r})\vec{e}_{z}$, in which $d_{x}(\vec{r})\vec{e}_{x}, d_{y}(\vec{r})\vec{e}_{y}, d_{z}(\vec{r})\vec{e}_{z}$ are all basis. Then we can rescale the order parameter with this substitution: $\vec{d}^{I}=v_{1}d_{x}\vec{e}_{x}+v_{2}d_{y}\vec{e}_{y}+v_{3}d_{z}\vec{e}_{z}$ with  $\alpha_{i}'|d_{i}|^2=\alpha'|\vec{d}^{I}_i|^2$ and $\alpha'$ equals to $\alpha_{i}'$ which corresponds to the highest transition temperature. $\alpha_{j}\ge 0$ for all $j=x,y,z$, that is to say, here we consider the condition $T\ge T_{c}$ since our goal is to get the transition temperature $T_{c}$ from effective free energy $f_{sc}$. The $\vec{d}^{I}$ can be expressed as:
\begin{equation}\label{di}
\vec{d}^{I}=\sqrt{\frac{\alpha_{x}'}{\alpha'}}d_{x}\vec{e}_{x}+\sqrt{\frac{\alpha_{y}'}{\alpha'}}d_{y}\vec{e}_{y}+\sqrt{\frac{\alpha_{z}'}{\alpha'}}d_{z}\vec{e}_{z},
\end{equation}
with $\alpha'=\alpha_{i}'$. Then the effective superconducting free energy Eq.\ref{fsc-1} changes into:
\begin{equation}
\alpha'|\vec{d}^{I}|^2+K_{2z}'m_{z0}p_{z}^{I}+\beta_{1jk}'|d_{j}^{I}|^2|d_{k}^{I}|^2+\beta_{2z}'(p_{z}^{I})^2+\beta_{3\nu z}' p_{z}^{I}|d_{\nu}^{I}|^2,
\end{equation}
where  $K_{2z}'=K_{2z}\frac{\sqrt{\alpha_{x}' \alpha_{y}'}}{\alpha'}$,  $\beta_{1jk}'=\frac{(\alpha')^{2}}{\alpha_{j}'\alpha_{k}'}\beta_{1jk}$,  $\beta_{2z}'=\beta_{2z}\frac{(\alpha')^{2}}{\alpha_{x}'\alpha_{y}'}$, $\beta_{3\nu z}'=\beta_{3\nu z}\frac{(\alpha')}{\sqrt{\alpha_{x}'\alpha_{y}'}}\frac{\alpha'}{\alpha_{\nu}'}$ (here, don't sum for the same index). In addition, we also omit the temperature dependence of the $K_{2z}'$ and $\beta'$s to derive the $T_{c}$ conveniently.

\section{III. Non-unitary Superconducting state}\label{app1}
We mainly focus on the superconducting critical temperature $T_{c}$, thus we can omit some quartic terms in effective free energy to get it approximatively:
\begin{equation}\label{a1}
f_{SC}=\alpha'|\vec{d}^{I}|^2+K_{2z}'p_{z}^{I}m_{z0}+\beta |\vec{d}^{I}|^4,
\end{equation}
By minimizing the free energy and using this equation: $\sum_{l}{d_{l} \frac{\partial f_{SC}}{\partial (d^{I})^{*}_{l}}}=0$, we can get:
\begin{equation}\label{a2}
\alpha' |\vec{d}^{I}|^2+2\beta |\vec{d}^{I}|^4+K_{2z}'m_{z0}(i \vec{d}^{I}\times\vec{(d^{I})^{*}})_{z}=0.
\end{equation}
For co-representation A:
substituting $\vec{d}^{I}=(d_{1}e^{i \theta},d_{2}e^{i \phi},d_{3})$ into Eq.\ref{a2} gives this equation:
\begin{equation}\label{a3}
\begin{split}
\alpha' (d_{1}^2+d_{2}^2+d_{3}^2)+2\beta (d_{1}^2+d_{2}^2+d_{3}^2)^2\\+2K_{2z}'m_{z0}d_{1}d_{2}sin({\phi-\theta})=0.
\end{split}
\end{equation}
For each point in real space, the amplitude of the order parameter $\vec{d}$ can be regarded as a function of coordinate of this point $\vec{r}$. To simplify this problem, and show the physics clearly, we take:
\begin{equation}\label{a4}
(d_{1},d_{2},d_{3})=d_{0}(\vec{r})(sin{x}cos{y},sin{x}sin{y},cos{x}),
\end{equation}
and obtain from Eq.\ref{a3}:
\begin{equation}\label{a5}
d_{0}(\vec{r})^2=\frac{-(\alpha'+K_{2z}' m_{z0} sin^2{x} sin{2y} sin{(\phi-\theta)})}{2\beta}.
\end{equation}
Where $\beta>0$. To get the highest superconducting critical temperature, in Eq.\ref{a5}, for $K_{2z}'>0$, we choose: $sin^2{x}=1,sin{2y}=1,sin({\phi-\theta})=-1$. So the d-vector is $\frac{d_{0}(\vec{r})}{\sqrt{2}}(1,-i,0)$. In the same way, for $K_{2z}'<0$, the d-vector is $\frac{d_{0}(\vec{r})}{\sqrt{2}}(1,i,0)$.
For co-representation B, the same method as in co-representation A can be used and the same result can be obtained.

\section{IV. Magnetic transitions in $URhGe$ ($UCoGe$)}\label{app3}

We start from the free energy Eq.(7)  about the magnetic moment in the main text:
\begin{equation}
f_{m}=-\frac{h_y^2}{4A_{1y}}+A^{'}_{1z} m_z^2+B^{'}_{z}m_{z}^4+C^{'}_{z}m_{z}^6.
\end{equation}
There are two non-zero local minima in this free energy which satisfy this equation:
\begin{equation}\label{eq1}
2A'_{1z} m_z^2+4B^{'}_{z}m_{z}^4+6C^{'}_{z}m_{z}^6=0.
\end{equation}
The solution of the Eq.\ref{eq1} is: $m_{z0}^2=\frac{-B^{'}_{z}+\sqrt{B^{'2}_{z}-3A^{'}_{1z}C^{'}_{z}}}{3C^{'}_{z}}$. The local minimum will be broken at this condition: $3C^{'}_{z} A^{'}_{1z}=B^{'2}_{z} $, namely, $m_{z e}^2=-\frac{B^{'}_{z}}{3C^{'}_{z}}\Leftrightarrow m_{z e}^4=\frac{A_{1z}'}{3C^{'}_{z}}$ ($A^{'}_{1z},C^{'}_{z}>0;B^{'}_{z}<0$). The metastable broken line is shown in the Fig.1, and corresponds to the Fig.2(d). Moreover, the first order transition line is determined by these equations:

\begin{equation}
\begin{split}
f_{para}=-\frac{h_y^2}{4A_{1y}}=f_{m}\Leftrightarrow A^{'}_{1z} m_z^2+B^{'}_{z}m_{z}^4+C^{'}_{z}m_{z}^6=0\\2A^{'}_{1z} m_z^2+4B^{'}_{z}m_{z}^4+6C^{'}_{z}m_{z}^6=0.
\end{split}
\end{equation}

By solving these equations, the condition of the first-order transition can be derived: $m_{z1}^2=-\frac{B'_{z}}{2C'_{z}}$, which is shown in Fig.1 and corresponds to the Fig.2 (b) in the main text. The breaking of the metastable state can also be understood via the disappearing of the extra minimum of the free energy $f_m$. Therefore, it is the point of the vanishing of both the first  and second order derivatives of $f_m$.

\section{V. Magnetic transitions in $UTe_{2}$}\label{app2}
\begin{figure}[]
    \centering
    \includegraphics[width=1\linewidth]{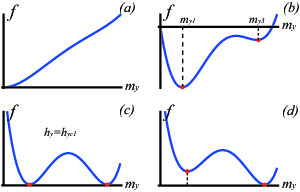}
    \caption{First-order transition in $UTe_2$, from (a) to (d), the magnetic field increases gradually. (a) free energy($f_{m}$)-magnetic momentum($m_{y}$) in zero magnetic field. (b)$f_{m}-m_{y}$ on the left side of first-order transition line. (c)$f_{m}-m_{y}$ on the first-order transition line. (d)$f_{m}-m_{y}$ on the right side of the first-order transition}
    \label{fig:a1}
\end{figure}

In the main text, we showed the free energy Eq.(14). Here we will talk about the first-order transition in $UTe_{2}$ in detail. If there are three real solutions for the Eq.\ref{app2-b1}, which are assumed as $m_{y1}< m_{y2} <m_{y3}$. $f_{m}(m_{y1})$ and $f_{m}(m_{y3})$ are the local minima, while $f_{m}(m_{y2})$ is the local maximum. As shown in Fig.\ref{fig:a1}, when the magnetic field increases, the minimum of the free energy will change from the $f_{m}(m_{y1})$ to the $f_{m}(m_{y3})$. This is the jump of the $m_{y0}$ at the critical field $h_{yc1}$. The critical magnetic field satisfies these equations:
\begin{align}
\frac{\partial{f_{m}}}{\partial{m_{y}}}=2a_{y}m_{y}+2b_{y}m_{y}^{3}-3c_{y}m_{y}^{2}-\mu_{1} h_{y}=0\label{app2-b1}\\
f(m_{y1})=f(m_{y3}).\label{app2-b2}
\end{align}

The relationship of the coefficients and the solutions is given by:
\begin{equation}\label{app2-b3}
\begin{split}
m_{y1}+m_{y2}+m_{y3}=\frac{3c_{y}}{2b_{y}}\\
m_{y1}m_{y2}+m_{y2}m_{y3}+m_{y3}m_{y1}=\frac{a_{y}}{b_{y}}\\
m_{y1}m_{y2}m_{y3}=\frac{\mu_{1} h_{y}}{2b_{y}}.
\end{split}
\end{equation}
From Eq.\ref{app2-b2}, this condition $m_{y1}+m_{y3}=2m_{y2}$ can be derived. By substituting it into the Eq.\ref{app2-b3}, we can get the equations about $m_{y1}$ and $m_{y3}$:
\begin{equation}
\begin{split}
m_{y1}+m_{y3}=\frac{c_{y}}{b_{y}}\\
m_{y1}m_{y3}=\frac{\mu_{1} h_{y}}{c_{y}},
\end{split}
\end{equation}
and the critical magnetic field: $h_{yc1}=\frac{c_{y}}{\mu_{1}}(\frac{a_{y}}{b_{y}}-\frac{c^{2}_{y}}{2b^2_{y}})$, which is corresponding to the magnetic field in the Fig.\ref{fig:a1} (c). The jump of the magnetic momentum is $\Delta m_{y0}=\sqrt{(\frac{c_{y}}{b_{y}})^2-\frac{4\mu_{1} h_{yc1}}{c_{y}}}$.

However, when the magnetic field is small, the local minima and local maximum don't appear, as shown in Fig.\ref{fig:a1} (a). So we can derive a critical magnetic field for the appearance of the local minima and local maximum. The critical magnetic field $h_{yc2}$ satisfies these equations:
\begin{equation}
\begin{split}
2a_{y}m_{y}+2b_{y}m_{y}^{3}-3c_{y}m_{y}^{2}-\mu_{1} h_{y}=0 \\2a_{y}+6b_{y}m_{y}^{2}-6c_{y}m_{y}=0.
\end{split}
\end{equation}
The critical magnetic field is $h_{yc2}=\frac{\Delta_{y} (8 a_{y} b_{y}-c_{y}\Delta_{y} )}{36 b^2_{y} \mu_{1}}$, where $\Delta_{y}=3 c_{y}+\sqrt{9 c_{y}^2-12 a_{y} b_{y}}$.

\section{VI. Two jumps in specific heat}
We start from the Eq.(2) in the main text, to deal with this problem conveniently and reveal the physics clearly, we simplify the free energy Eq.(2) and rewrite it as:
\begin{equation}\label{scfs}
f_{sc}^{I}=\alpha'_{i}|d_{i}|^2+K_{2z}m_{z0}p_{z}+\beta_{1i}|d_{i}|^4.
\end{equation}
Order parameters can be gotten by: $\frac{\partial f_{sc}^{I}}{\partial d_{i}^{*}}=0$:
\begin{equation}\label{ds}
\begin{split}
x: \alpha_{x}'d_{x}-i K_{2z}m_{z0}d_{y}+2\beta_{1x}d_{x}|d_{x}|^2=0;\\
y: \alpha_{y}'d_{y}+i K_{2z}m_{z0}d_{x}+2\beta_{1y}d_{y}|d_{y}|^2=0;\\
z: \alpha_{z}'d_{z}+2\beta_{1z}d_{z}|d_{z}|^2=0.
\end{split}
\end{equation}
To show the effect of the $K_{2z}$ coupling term clearly, we assume the transition temperatures without $K_{2z}$ coupling terms are the same. Therefore, we can simplify the Eq.\ref{ds} as:
\begin{equation}
\begin{split}
x: \alpha'd_{x}-i K_{2z}m_{z0}d_{y}+2\beta d_{x}|d_{x}|^2=0;\\
y: \alpha'd_{y}+i K_{2z}m_{z0}d_{x}+2\beta d_{y}|d_{y}|^2=0;\\
z: \alpha'd_{z}+2\beta_{1z}d_{z}|d_{z}|^2=0.
\end{split}
\end{equation}
Solving these equations and taking the highest temperature for the $x$ and $y$ components of the order parameter $\vec{d}$ give two transition temperatures:
\begin{equation}
\begin{split}
T_{cz}=T'\equiv T_{1};\\
T_{cx}=T_{cy}=T'+\frac{K_{2z}m_{z0}}{\alpha_{0}'}\equiv T_{2}.
\end{split}
\end{equation}
Thus we can rewrite the components of the order parameter as:
\begin{equation}
\begin{split}
|d_{x}|^2=-\frac{\alpha_{0}'(T-T_{2})}{2\beta};\\
|d_{y}|^2=-\frac{\alpha_{0}'(T-T_{2})}{2\beta};\\
|d_{z}|^2=-\frac{\alpha'_{0}(T-T_{1})}{2\beta_{1z}}.
\end{split}
\end{equation}
With free energy Eq.\ref{scfs},  the entropy can be expressed as:
\begin{equation}\label{sapp}
    \begin{split}
        S=S_{n}, T>T_{2}\\
        S=S_{n}-\alpha_{x0}|d_{x}|^2-\alpha_{y0}|d_{y}|^2, T_{1}<T<T_{2}\\
        S=S_{n}-\alpha_{x0}|d_{x}|^2-\alpha_{y0}|d_{y}|^2-\alpha_{z0}|d_{z}|, T<T_{2}.
    \end{split}
\end{equation}
Where the $S_{n}$ is the entropy of the normal state, $\alpha_{i0} (i=x,y,z)$ are positive constants. Thus the slope of the entropy will change when $T=T_{1} $ and $T_{2}$.

The specific heat can be derived from Eq.\ref{sapp}:
\begin{equation}
\begin{split}
    C=C_{n}, T>T_{2}\\
    C=C_{n}+T \frac{\alpha_{x0}\alpha_{0}'}{2\beta}+T \frac{\alpha_{y0}\alpha_{0}'}{2\beta}, T_{1}<T<T_{2}\\
    C=C_{n}+T \frac{\alpha_{x0}\alpha_{0}'}{2\beta}+T \frac{\alpha_{y0}\alpha_{0}'}{2\beta}+T \frac{\alpha_{z0} \alpha_{0}'}{2\beta_{1z}}, T<T_{2}.
\end{split}
\end{equation}
Thus the jumps of the specific heat can be expressed as:

\begin{equation}
    \begin{split}
    \Delta C_{1}=T_{2}(\frac{\alpha_{x0}\alpha_{0}'}{2\beta}+ \frac{\alpha_{y0}\alpha_{0}'}{2\beta})\\
    \Delta C_{2}=T_{1} \frac{\alpha_{z0} \alpha_{0}'}{2\beta_{1z}}\\
    \end{split}
\end{equation}

Clearly, the spefici heat jump over the transition temperature do not dependent on the magnetic field and can be used to extract the phenomenological coefficients from the experiments.
\end{appendix}

\end{document}